\documentclass[10pt,leqno]{amsart}

\usepackage[english]{babel}
\usepackage[T1]{fontenc}
\usepackage[utf8]{inputenc}
\usepackage{graphicx}
\usepackage{indentfirst,csquotes}
\usepackage{amssymb,amsthm,amsmath}
\usepackage{xcolor}
\usepackage{paralist}
\usepackage{hyperref}
\usepackage{fancyhdr}
\usepackage{etoolbox}
\usepackage[numbers]{natbib}
\usepackage{booktabs}
\usepackage{array}

\hypersetup{colorlinks=true, linkcolor=black, filecolor=black, urlcolor=black, citecolor=black}

\begin{document}

\title[SWAN]{Training a generalizable diffusion model for seismic data processing using a large-scale open-source waveform dataset}
\author{Xinyue Gong, Sergey Fomel, and Yangkang Chen}
\date{January 2026}
\maketitle

\begin{abstract}
We introduce the Seismic Waveforms dataset for Automatic Neural-network processing (SWAN), a comprehensive and standardized benchmark designed to advance data-driven seismic signal processing. SWAN aggregates diverse synthetic and real seismic waveforms spanning a wide range of geological structures, noise conditions, propagation environments, and acquisition geometries, providing a unified foundation for training highly generalizable models. Leveraging this dataset, we develop and evaluate a conditionally constrained residual diffusion model for core seismic processing tasks, focusing on missing-trace reconstruction. Extensive experiments demonstrate that diffusion models trained on SWAN achieve state-of-the-art performance across heterogeneous testing scenarios, outperforming leading deep-learning and physics-based baselines on both synthetic benchmarks and field data examples. The results highlight SWAN's value as both a scalable training corpus and a rigorous evaluation framework, and illustrate the strong potential of diffusion-based architectures for robust, generalizable seismic data processing.
\end{abstract}

\section{Introduction}

Deep learning (DL) has emerged as a transformative paradigm in seismic data processing and imaging, leading to substantial improvements across core tasks such as noise attenuation \citep{yang2021deep, saad2020deep, Li2024_RankReduction}, missing-trace reconstruction \citep{Gong2023_CSDNN, saad2023unsupervised}, deblending \citep{Li2025_OTGDeblend} and full-wave inversion (FWI) \citep{saad2024siamesefwi}. Modern neural architectures, ranging from convolutional neural networks (CNNs) to transformer-based models \citep{Li2025_FaultVitNet} and diffusion models \citep{Gong2025b_Dual}, have demonstrated strong capability in learning the spatiotemporal complexity of seismic wavefields.

Although these developments are encouraging, progress in DL for seismic processing remains constrained by a persistent data bottleneck \citep{reichstein2019deep}. Publicly available seismic datasets are limited in scale, heterogeneous in format, and often not provided in an AI-ready form suitable for modern end-to-end learning pipelines \citep{wu2023sensing, sheng2025seismic}. Consistently preprocessed patch-level wavefield datasets are particularly scarce, even though they are essential for training models that can operate reliably across surveys, acquisition geometries, and noise environments.

This limitation has become increasingly evident with the emergence of geophysical foundation models. The Seismic Foundation Model (SFM) \citep{sheng2025seismic}, for example, aggregates millions of seismic images to pretrain large self-supervised backbones for interpretation-oriented tasks. While such efforts demonstrate the potential of large-scale pretraining, they also expose a clear gap. The community lacks an accessible, standardized wavefield dataset specifically designed to support low-level processing tasks such as denoising, interpolation, and missing-trace reconstruction. Existing foundation-model datasets focus primarily on labels or structural interpretation targets rather than the raw wavefield representations needed for signal-processing workflows. As a result, reproducibility and fair comparison across DL-based seismic processing methods remain difficult.

Beyond dataset scale, diversity, and accessibility, current public resources also suffer from inconsistent preprocessing practices. Different surveys may adopt incompatible normalization strategies, patch sizes, noise handling, or coordinate conventions \citep{guo2025cross, sheng2025seismic}. These inconsistencies hinder cross-survey generalization and contribute to a long-standing challenge in seismic ML, where models succeed on narrowly defined datasets but fail to transfer across geological settings or acquisition conditions \citep{wu2023sensing}.

To address these limitations, we introduce the Seismic Waveforms Dataset for Automatic Neural-network processing (SWAN), a unified and AI-ready seismic patch corpus designed specifically for DL-based wavefield processing. SWAN contains 537,373 non-overlapping 128$\times$128 patches obtained from 20 synthetic and real surveys. Synthetic datasets contribute approximately 74.4\% of all patches, while real field datasets contribute the remaining 25.6\%. This combination provides both physics-consistent ground truth and the geological variability necessary for generalizable learning. Each patch is accompanied by comprehensive metadata describing acquisition geometry, normalization factors, spatial context, and quality indicators, enabling transparent data filtering and complete traceability.

SWAN offers several contributions that distinguish it from existing seismic datasets. First, it is explicitly designed for wavefield-level processing rather than structural interpretation, making it directly applicable to seismic reconstruction, denoising, and acquisition recovery \citep{yang2019deep}. Second, it functions as a foundation dataset that supports model training, benchmarking, and cross-survey evaluation across diverse DL architectures, including CNNs, transformers, and diffusion models. Third, it integrates synthetic and real data in a unified format, which narrows the domain gap between numerical simulations and field acquisition conditions and enables more robust and transferable learning \citep{guo2025cross}.

As seismic research continues to adopt data-centric AI practices \citep{schneider2024foundation}, SWAN provides a standardized and extensible platform for developing and evaluating DL-based seismic processing methods. By unifying data formatting, normalization, patch extraction, and metadata conventions, SWAN improves reproducibility, facilitates fair comparisons, and enables the broader community to build upon a common seismic data backbone \citep{tenopir2018research}. We expect SWAN to support both academic research and industrial applications, particularly in areas such as interpolation, noise suppression, and generative modeling for seismic reconstruction.

\section{SWAN Dataset}

To support large-scale and generalizable DL research for seismic wavefield processing, we construct the SWAN dataset. SWAN is a unified, AI-ready collection of seismic patches extracted from a wide range of synthetic and real surveys. The dataset addresses a longstanding bottleneck in seismic machine learning, namely the lack of standardized patch-level data that consistently represent both prestack and poststack wavefields. This section outlines the overall data processing workflow and describes the composition of the four major dataset categories.

\subsection{Data Processing Pipeline}

\begin{figure}[!t]
    \centering
    \includegraphics[width=\linewidth]{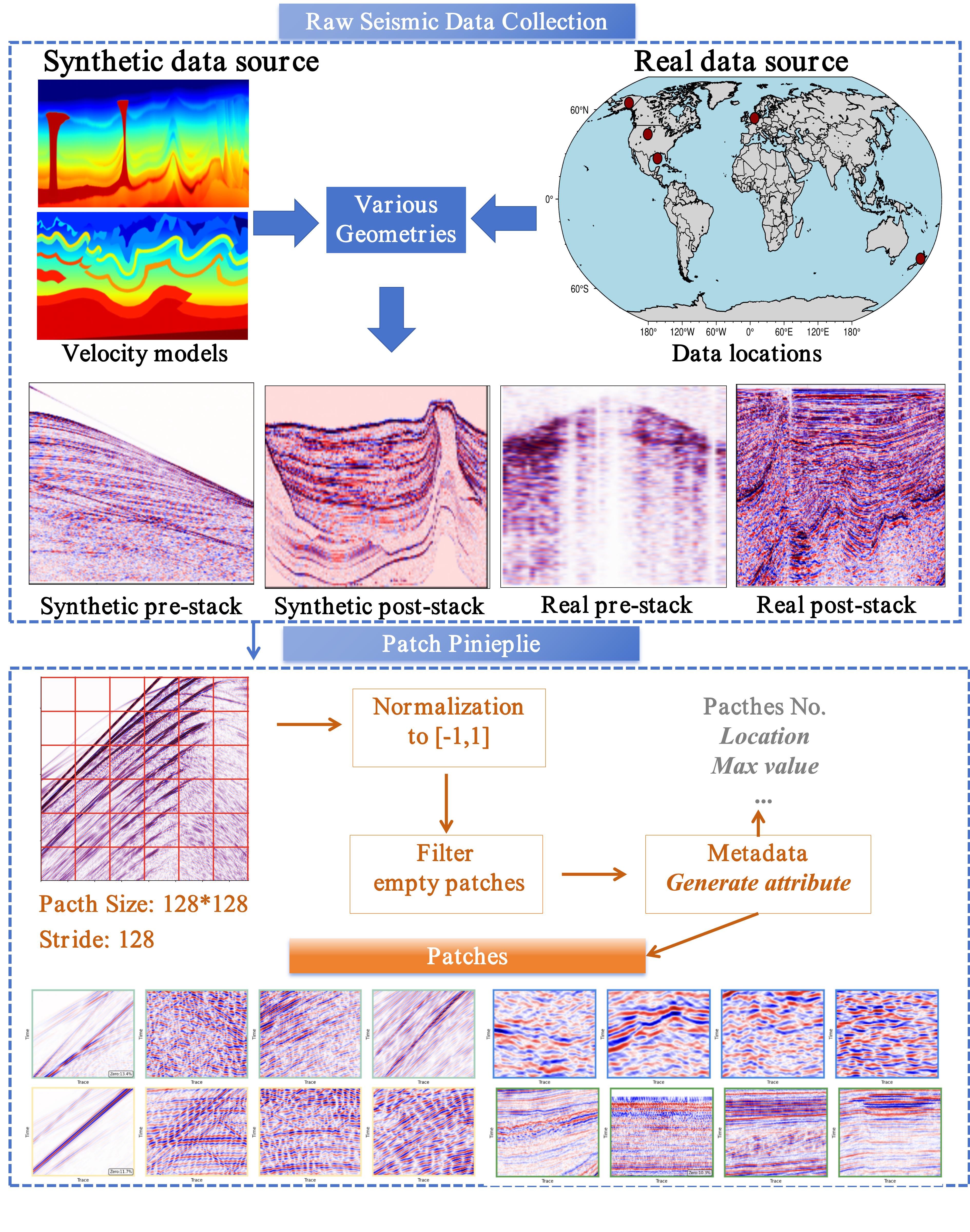}
    \caption{Overview of the SWAN data processing pipeline, including synthetic and real data sources, patch extraction, normalization, quality filtering, and metadata generation.}
    \label{fig:swan_pipeline}
\end{figure}

The SWAN dataset is produced through a unified workflow that standardizes diverse seismic surveys into consistently formatted $128 \times 128$ wavefield patches. The complete workflow is illustrated in Fig.~\ref{fig:swan_pipeline}. Synthetic surveys originate from benchmark velocity models, including BP 1994, BP 2004, BP 2007 TTI, BP 2.5D, Marmousi, Pluto, SEAM Phase I, and Amoco. Real surveys come from several major geological regions such as the Taranaki Basin in New Zealand, the North Sea F3 block, the Gulf of Mexico, Alaska, and Wyoming. These datasets cover both marine and land environments and include shot gathers as well as migrated sections.

All seismic volumes are represented as 2D wavefields and partitioned into fixed-size patches using a $128 \times 128$ sliding window with a stride of 128 samples. Each patch is normalized by its maximum absolute amplitude, resulting in values within $[-1, 1]$. This step eliminates the need for survey-specific scaling and allows patches from heterogeneous surveys to be used directly in DL workflows.

Quality control is performed automatically. For most datasets, patches containing more than 90\% zero values are removed because such tiles typically originate from edge padding, inactive traces, or empty recording windows. For several datasets with known acquisition or preprocessing characteristics, this threshold is adjusted to better preserve valid wavefield information. This flexible approach maintains dataset consistency while retaining meaningful seismic signals.

Each retained patch is stored together with metadata describing the survey name, patch position, time, and trace indices, normalization factors, and quality indicators such as zero ratio. These metadata support reproducibility, survey-level filtering, reconstruction of spatial context, and flexible dataset selection for DL model training. In total, SWAN contains 537,373 patches extracted from 20 datasets.

\subsection{Dataset Composition}

\begin{figure*}[t]
    \centering
    \includegraphics[width=\linewidth]{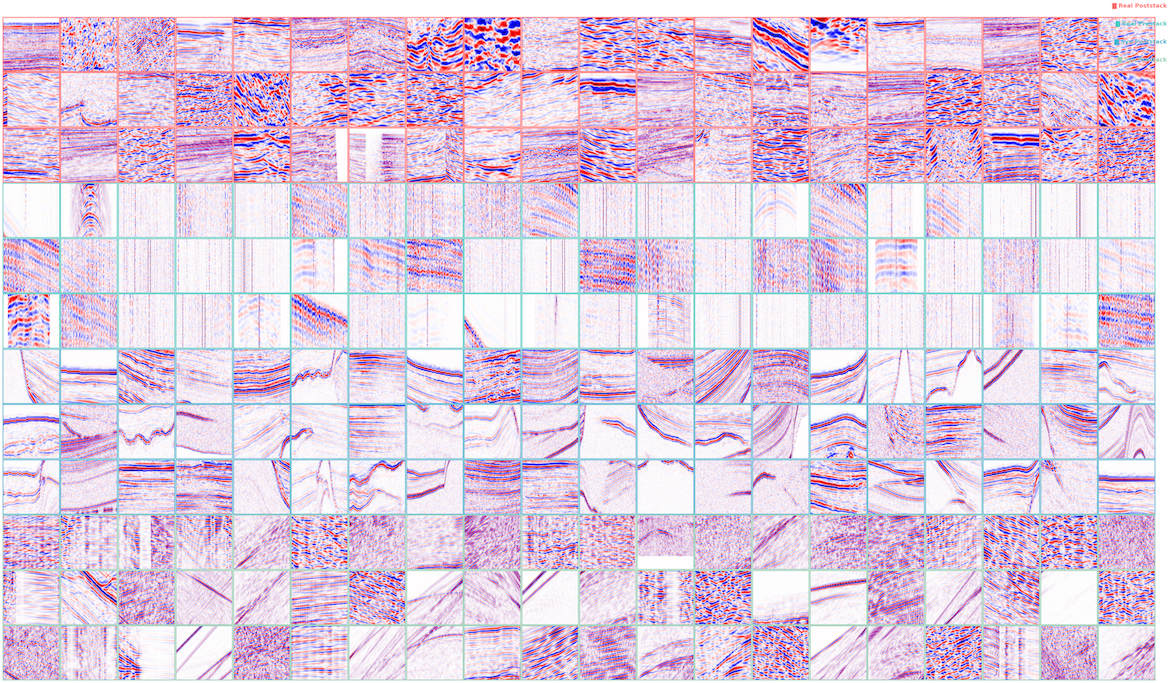}
    \caption{Representative $128\times128$ patches sampled from the four SWAN categories. Each group of three rows corresponds to one data type and is outlined using a distinct border color: real poststack (red, rows 1--3), real prestack (teal, rows 4--6), synthetic poststack (blue, rows 7--9), and synthetic prestack (green, rows 10--12).}
    \label{fig:samples_4_types}
\end{figure*}

The SWAN dataset integrates 20 constituent datasets grouped into four major categories. Synthetic datasets contribute approximately 74.4\% of all patches, while real surveys contribute 25.6\%. Although the distribution is imbalanced, additional field datasets will be incorporated in future releases to expand real-data diversity. Representative examples from each category appear in Fig.~\ref{fig:samples_4_types}.

\begin{table*}[t]
\caption{Detailed summary of the SWAN dataset collections.}
\label{tab:swan_detailed}
\centering
\setlength{\tabcolsep}{2.5pt}
\renewcommand{\arraystretch}{1.5}
{\footnotesize
\begin{tabular}{lccccp{3.6cm}}
\toprule
\textbf{Dataset} &
\textbf{Files} &
\textbf{Patches} &
\textbf{Dim (samples $\times$ traces)} &
\textbf{dt (ms)} &
\textbf{Region / Source} \\
\midrule
\multicolumn{6}{c}{\textbf{Synthetic Prestack}} \\
\midrule
1994bp & 876 & 137,559 & 1152 $\times$ 3008 & 5.4 & BP Benchmark \\
1994bp\_mig & 278 & 7,557 & 2000 $\times$ 240--480 & 4.0 & BP Migration Test \\
1997bp\_2.5d & 1 & 1,991 & 384 $\times$ 98,560 & 9.9 & BP 2.5D Model \\
2004bp\_velocity & 1,354 & 117,638 & 2001 $\times$ 1201 & 6.0 & BP Velocity Model \\
2007bp\_tti & 1,641 & 49,599 & 1151 $\times$ 800 & 8.0 & BP TTI Model \\
Amoco & 385 & 1,086 & 384 $\times$ 256 & 9.9 & AMOCO \\
Marmousi & 1 & 982 & 1500 $\times$ 256 & 4.0 & SMAART JV \\
Pluto & 694 & 9,081 & 1126 $\times$ 350 & 8.0 & SMAART JV \\
\midrule
\multicolumn{6}{c}{\textbf{Synthetic Poststack}} \\
\midrule
SEAM\_1 & 1002 & 38,596 & 851 $\times$ 1169 & 4.0 & SEAM Phase I \\
SEAM\_2 & 1169 & 35,927 & 851 $\times$ 1002 & 4.0 & SEAM Phase I \\
\midrule
\multicolumn{6}{c}{\textbf{Real Prestack}} \\
\midrule
Alaska & 58 & 1,300 & 3000 $\times$ 95 & 2.0 & USGS NPRA Alaska \\
OZ~Yilmaz & 40 & 471 & 975--2535 $\times$ 24--216 & 4.0 & SEG Textbook \\
Stratton & 102 & 5,198 & 3000 $\times$ 328--652 & 2.0 & OSG Texas \\
\midrule
\multicolumn{6}{c}{\textbf{Real Poststack}} \\
\midrule
KERRY & 735 & 11,597 & 1252 $\times$ 287 & 4.0 & New Zealand \\
WAIHAPA & 305 & 5,795 & 2501 $\times$ 227 & 2.0 & New Zealand \\
WAIPUKU & 312 & 4,107 & 2001 $\times$ 148 & 2.0 & New Zealand \\
BOEM & 1 & 7,302 & 450 $\times$ 311,640 & 10.0 & Gulf of Mexico \\
F3 & 651 & 13,290 & 462 $\times$ 951 & 4.0 & Dutch North Sea \\
KAHU & 1 & 85,437 & 1500 $\times$ 994,230 & 4.0 & New Zealand \\
Teapot~Dome & 1 & 2,860 & 1501 $\times$ 33,286 & 2.0 & RMOTC Wyoming \\
\bottomrule
\end{tabular}
}
\end{table*}

\textbf{Synthetic Prestack.}
This category includes eight modeling benchmarks, such as BP 1994, BP 2.5D, BP 2004 velocity, BP migration, BP 2007 TTI, Marmousi, Pluto, and Amoco. These datasets exhibit clean reflection events, diffractions, anisotropic wavefronts, and long-offset kinematic patterns. Their controlled nature makes them particularly suitable for evaluating DL-based interpolation and generative reconstruction. Representative examples appear in the green-bordered rows of Fig.~\ref{fig:samples_4_types}.

\textbf{Synthetic Poststack.}
Two SEAM Phase I datasets extracted along the inline and crossline directions belong to this category. They capture structurally complex deepwater geology, including salt bodies, sharply contrasting reflectors, and faulted sedimentary units. These characteristics provide a rich test bed for structure-preserving denoising and post-migration enhancement. Representative examples appear in the blue-bordered rows of Fig.~\ref{fig:samples_4_types}.

\textbf{Real Prestack.}
Real prestack datasets include Stratton 3D, the USGS Alaska line, and the Oz marine survey. They contain acquisition challenges such as statics, ground roll, swell noise, irregular offsets, and nonstationary amplitude decay. These properties make them essential for assessing the robustness and generalization of DL-based reconstruction. Representative examples appear in the teal-bordered rows of Fig.~\ref{fig:samples_4_types}.

\textbf{Real Poststack.}
This category includes a large collection of migrated 2D and 3D surveys, including F3 (North Sea), Kerry, Waihapa, Waipuku, BOEM Gulf of Mexico, Kahu, and Teapot Dome. These datasets exhibit diverse structural features such as faults, channels, dipping reflectors, and unconformities. They provide geologically meaningful patterns for evaluating structural fidelity in DL-based processing. Representative examples appear in the red-bordered rows of Fig.~\ref{fig:samples_4_types}.

\section{Methodology}

\subsection{Diffusion Models}

Diffusion models approximate complex data distributions by gradually corrupting a clean signal and learning a neural network that inverts this transformation. Let $\mathbf{x}_0$ denote a clean seismic patch. The forward diffusion process applies a sequence of Gaussian perturbations:

\begin{equation}
q(\mathbf{x}_t \mid \mathbf{x}_{t-1})
= \mathcal{N}\!\left(\sqrt{\alpha_t}\mathbf{x}_{t-1}, (1-\alpha_t)\mathbf{I}\right),
\label{eq:ddpm_forward}
\end{equation}
with the closed-form marginal

\begin{equation}
\begin{aligned}
\mathbf{x}_t
&= \sqrt{\bar{\alpha}_t}\,\mathbf{x}_0
+ \sqrt{1-\bar{\alpha}_t}\,\boldsymbol{\epsilon}, \\
\boldsymbol{\epsilon}
&\sim \mathcal{N}(0, \mathbf{I}),
\end{aligned}
\label{eq:ddpm_marginal}
\end{equation}
where $\bar{\alpha}_t = \prod_{s=1}^{t}\alpha_s$. A neural network estimates the injected noise by minimizing

\begin{equation}
\mathcal{L}_{\mathrm{DDPM}}
= \mathbb{E}\!\left[
\|\boldsymbol{\epsilon}
- \boldsymbol{\epsilon}_\theta(\mathbf{x}_t, t)\|^2
\right].
\label{eq:ddpm_loss}
\end{equation}

This classical formulation is effective for image generation, yet it relies on isotropic Gaussian corruption and produces stochastic reverse trajectories. These properties conflict with the structured nature of seismic degradation, which is dominated by spatially coherent missing traces rather than random noise. Moreover, reversing diffusion from pure noise disregards the information present in the observed waveform and introduces sampling variance. These issues motivate a diffusion formulation that is more tightly aligned with seismic physics and reconstruction objectives.

\subsection{Residual-Guided Diffusion Model (RGDM)}

\begin{figure*}[t]
    \centering
    \includegraphics[width=\linewidth]{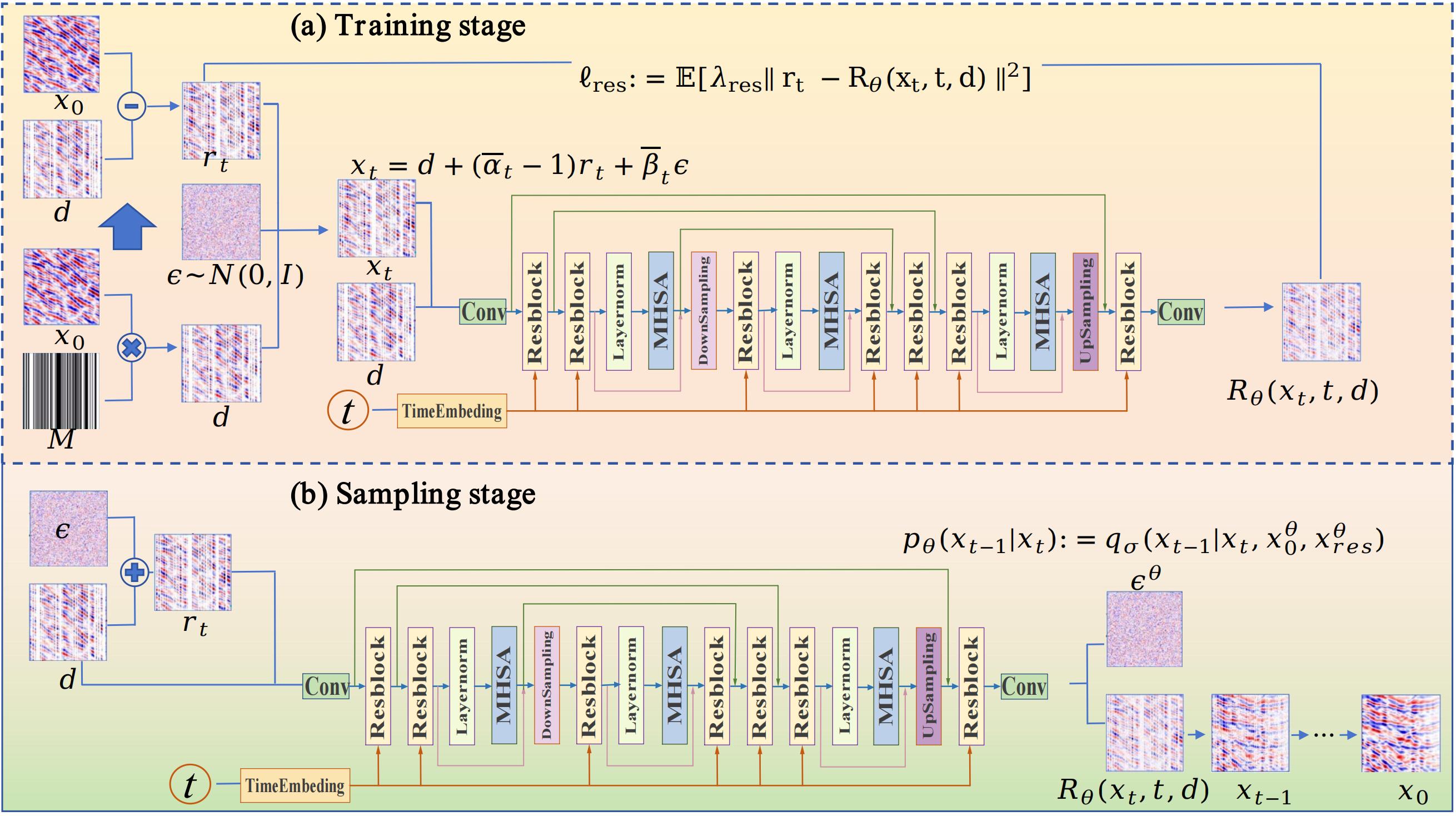}
    \caption{Residual-guided diffusion framework used for seismic reconstruction. The training stage (top) learns residual increments, while the sampling stage (bottom) applies deterministic reverse diffusion conditioned on the observed waveform.}
    \label{fig:rgdm_framework}
\end{figure*}

To address these limitations, we introduce the Residual-Guided Diffusion Model (RGDM). The key idea is to reformulate diffusion as a residual-correction process that remains anchored to the observed seismic waveform instead of drifting toward pure noise. RGDM models the latent evolution as a sequence of deterministic corrections that reflect the discrepancy between the observed and clean wavefields. The training and sampling workflows are shown in Fig.~\ref{fig:rgdm_framework}.

\subsubsection{Forward Process}

Let $\mathbf{d}=\mathbf{M}\odot\mathbf{x}_0$ denote the observed gather. RGDM initializes the terminal diffusion state using a mild Gaussian perturbation:

\begin{equation}
\mathbf{x}_T = \mathbf{d} + \sigma_T \boldsymbol{\epsilon}.
\label{eq:rgdm_init}
\end{equation}

The forward evolution then proceeds through residual increments:

\begin{equation}
\mathbf{x}_t
= \mathbf{x}_{t-1}
+ r_t
+ \beta_t \boldsymbol{\epsilon}_t,
\label{eq:rgdm_forward}
\end{equation}
where $r_t$ represents the deviation between $\mathbf{x}_{t-1}$ and the underlying clean manifold associated with $\mathbf{x}_0$. This formulation ensures that the latent trajectory remains centered on the observed data rather than diffusing toward noise. As a result, the forward process captures a physically meaningful correction pathway that reflects seismic acquisition effects.

\subsubsection{Reverse Process}

The reverse process predicts step-wise corrections that cancel the accumulated residuals. A U-Net-based rectifier $\mathcal{R}_\theta(\mathbf{x}_t, t, \mathbf{d})$ estimates

\begin{equation}
\hat{r}_t
= \mathcal{R}_\theta(\mathbf{x}_t, t, \mathbf{d}),
\label{eq:rgdm_residual}
\end{equation}
and the reverse update becomes

\begin{equation}
\mathbf{x}_{t-1}
= \mathbf{x}_t
- \hat{r}_t
- \frac{\beta_t^2}{\bar{\beta}_t}\hat{\boldsymbol{\epsilon}}_t
+ \eta_t \boldsymbol{\epsilon},
\label{eq:rgdm_reverse}
\end{equation}
with the induced noise estimate:

\begin{equation}
\hat{\boldsymbol{\epsilon}}_t
= \frac{\mathbf{x}_t
- \mathbf{d}
- \bar{r}_t(\theta)}{\bar{\beta}_t}.
\label{eq:rgdm_noise}
\end{equation}

The reverse dynamics remain deterministic when $\eta_t = 0$, producing a stable reconstruction trajectory that progressively removes artifacts while respecting the information contained in $\mathbf{d}$.

\subsubsection{Training Objective}

RGDM employs a U-Net backbone with temporal embeddings and multi-head self-attention to jointly process $\mathbf{x}_t$ and the observed waveform $\mathbf{d}$. Training directly matches the true and predicted residual increments:

\begin{equation}
\mathcal{L}_{\mathrm{RGDM}}
= \mathbb{E}\!\left[
\|r_t - \mathcal{R}_\theta(\mathbf{x}_t, t, \mathbf{d})\|^2
\right].
\label{eq:rgdm_loss}
\end{equation}

This objective enforces a physically interpretable correction path and encourages the model to learn how seismic acquisition patterns distort wavefields.

\section{Numerical Experiments}

To demonstrate the practical value and learning potential enabled by the SWAN dataset, we train our reconstruction model using 50,000 randomly selected patches from each data category. The experiments evaluate not only the reconstruction capability of the proposed method but also its generalization to datasets that never appear in the SWAN training subset. Several representative baselines are included for comparison, including projection onto convex sets (POCS) algorithm \citep{abma2006}, damped rank-reduction (DRR) \citep{yangkang2023drr}, and PySeistr \citep{pyseistr}. Unless otherwise specified, all experiments use 50\% irregular trace removal.

Most experiments follow a blind-test protocol. The test datasets include four synthetic benchmarks as well as several field examples, comprising a 2D hyperbolic gather, a 2D edge-structure gather, a 3D hyperbolic volume, a synthetic DAS gather, a Viking Graben 2D section, a SeanS3 3D volume, and a local field DAS segment. One non-blind experiment involving the 1997 BP dataset is also included to assess the impact on migration imaging.

\subsection{Example 1: Synthetic Hyperbolic Data}

\begin{figure*}[t]
    \centering
    \includegraphics[width=\textwidth]{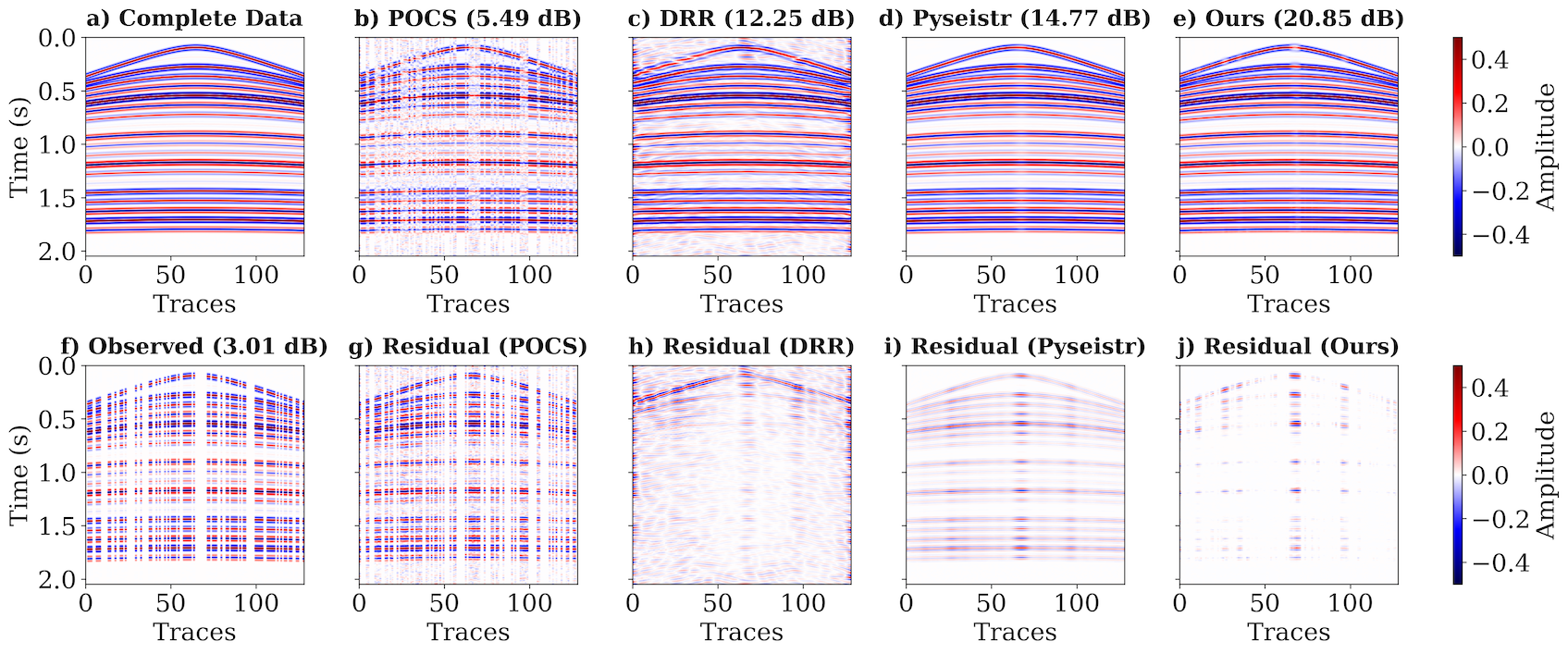}
    \caption{Example 1. Interpolation of a synthetic hyperbolic gather with 50\% irregular sampling. (a) Complete data. (b)--(e) Reconstruction results of POCS, DRR, PySeisTr, and the proposed method. (f) Observed gather with 50\% missing traces. (g)--(j) Corresponding residual panels.}
    \label{fig:syncoh2}
\end{figure*}

The first blind-test dataset was originally introduced by Zhou et al.~\citep{Zhou2020dictionary}. It contains a marine-style common-shot gather with several clean hyperbolic reflection events (Fig.~\ref{fig:syncoh2}a), making it a standard benchmark for interpolation under 50\% irregular sampling (Fig.~\ref{fig:syncoh2}f).

Figure~\ref{fig:syncoh2} shows the reconstruction results. POCS (Fig.~\ref{fig:syncoh2}b) recovers only low-wavenumber trends and introduces strong vertical streaks. DRR (Fig.~\ref{fig:syncoh2}c) improves reflector continuity but still leaves scattered incoherent energy. PySeisTr (Fig.~\ref{fig:syncoh2}d) suppresses a portion of the noise and better preserves event geometry, although substantial residual energy remains (Fig.~\ref{fig:syncoh2}i).

Our SWAN-trained model (Fig.~\ref{fig:syncoh2}e; SNR = 20.85\,dB) achieves the most coherent reconstruction. Reflection events exhibit continuous curvature, consistent phase alignment, and stable amplitudes across the entire gather. The residual panel (Fig.~\ref{fig:syncoh2}j) shows only weak and spatially isolated energy. These results indicate that the model has learned strong event-level priors from the SWAN dataset and generalizes effectively to unseen synthetic gathers.

\subsection{Example 2: Synthetic Edge-Structure Data}

The second blind-test dataset was introduced by Zhou et al.~\citep{Zhou2020online} for evaluating edge-preserving reconstruction methods. The gather contains strong lateral terminations and sharp discontinuities that challenge interpolation algorithms.

\begin{figure*}[t]
    \centering
    \includegraphics[width=\textwidth]{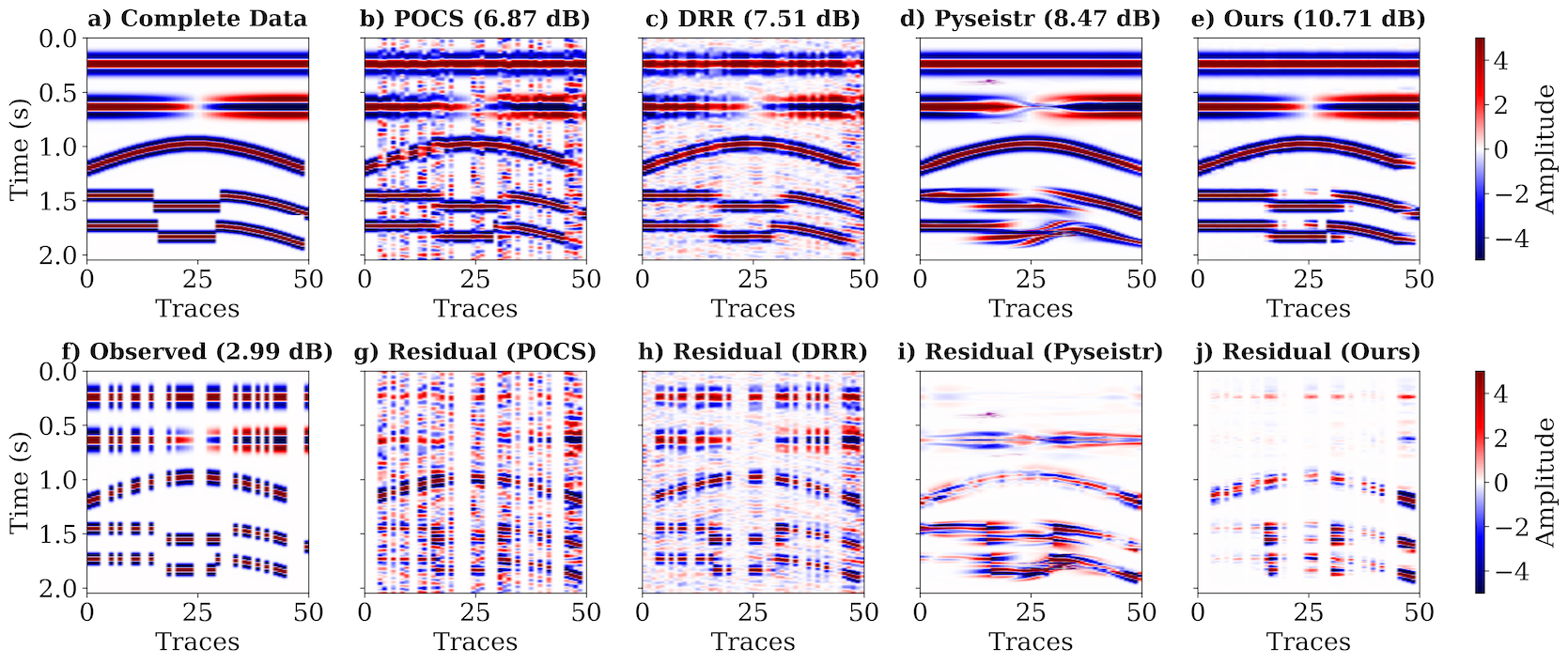}
    \caption{Example 2. Interpolation of a synthetic edge-structure gather with 50\% irregular sampling. (a) Complete data. (b)--(e) Reconstruction results of POCS, DRR, PySeisTr, and the proposed method. (f) Observed gather with 50\% missing traces. (g)--(j) Corresponding residual panels.}
    \label{fig:synedge2}
\end{figure*}

The observed data with 50\% trace removal (Fig.~\ref{fig:synedge2}f) exhibit fragmented edges and aliasing. POCS (Fig.~\ref{fig:synedge2}b) recovers primarily smooth background components. DRR (Fig.~\ref{fig:synedge2}c) improves reflector continuity but introduces noticeable artifacts in the missing regions. PySeisTr (Fig.~\ref{fig:synedge2}d) better reconstructs dipping events but leaves substantial residuals (Fig.~\ref{fig:synedge2}i).

Our model (Fig.~\ref{fig:synedge2}e; SNR = 10.71\,dB) accurately restores both continuous reflections and abrupt terminations. The residuals (Fig.~\ref{fig:synedge2}j) are weak and spatially compact, demonstrating that the SWAN-trained model captures geometric complexity even though such edge structures were not explicitly seen during training.

\subsection{Example 3: 3D Synthetic Hyperbolic Volume}

The third experiment uses a 3D synthetic hyperbolic volume introduced by Bai et al.~\citep{Bai2020}. The volume contains dipping and curved reflectors with strong spatial continuity, making it suitable for evaluating reconstruction in three dimensions.

\begin{figure*}[t]
    \centering
    \includegraphics[width=\textwidth]{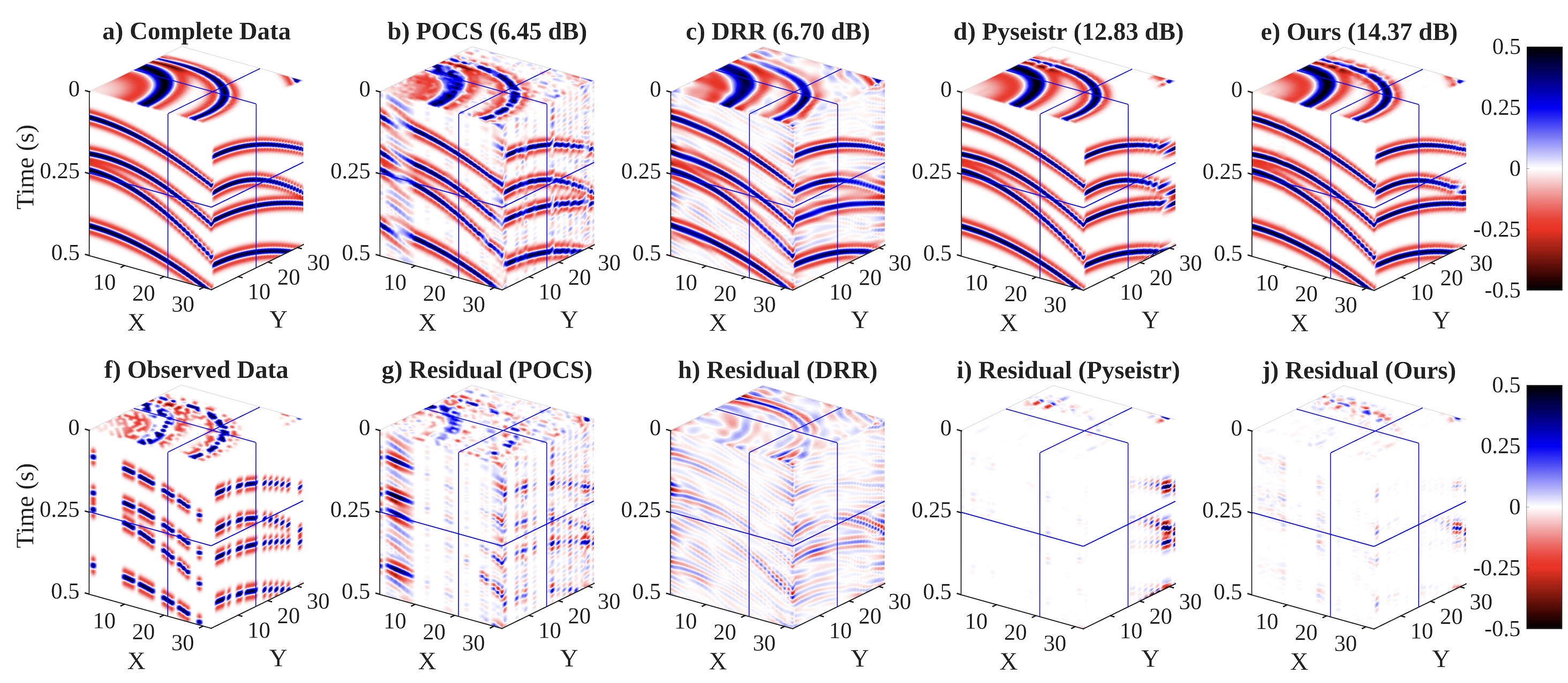}
    \caption{Example 3. Interpolation of a 3D synthetic hyperbolic volume with 50\% irregular sampling. (a) Complete data. (b)--(e) Results of POCS, DRR, PySeisTr, and the proposed model. (f) Observed data with 50\% missing traces. (g)--(j) Residual panels.}
    \label{fig:hyper4}
\end{figure*}

The observed data with 50\% missing traces (Fig.~\ref{fig:hyper4}f) exhibit severe disruption of 3D continuity. POCS (Fig.~\ref{fig:hyper4}b) and DRR (Fig.~\ref{fig:hyper4}c) partially restore structures but leave significant incoherent energy. PySeisTr (Fig.~\ref{fig:hyper4}d; 12.83\,dB) produces more consistent reflectors yet introduces leakage around complex curvature.

The proposed method (Fig.~\ref{fig:hyper4}e; 14.37\,dB) reconstructs the 3D structures with the highest fidelity. The residuals (Fig.~\ref{fig:hyper4}j) remain weak and spatially isolated, demonstrating that the model generalizes effectively to volumetric settings even though training uses only 2D patches.

\subsection{Example 4: Synthetic DAS Data}

The fourth experiment evaluates a synthetic DAS dataset introduced by Chen et al.~\citep{Chen2023DAS}. The gather contains dipping reflections typical of surface DAS acquisitions, and 50\% of the channels are removed.

\begin{figure*}[t]
    \centering
    \includegraphics[width=\textwidth]{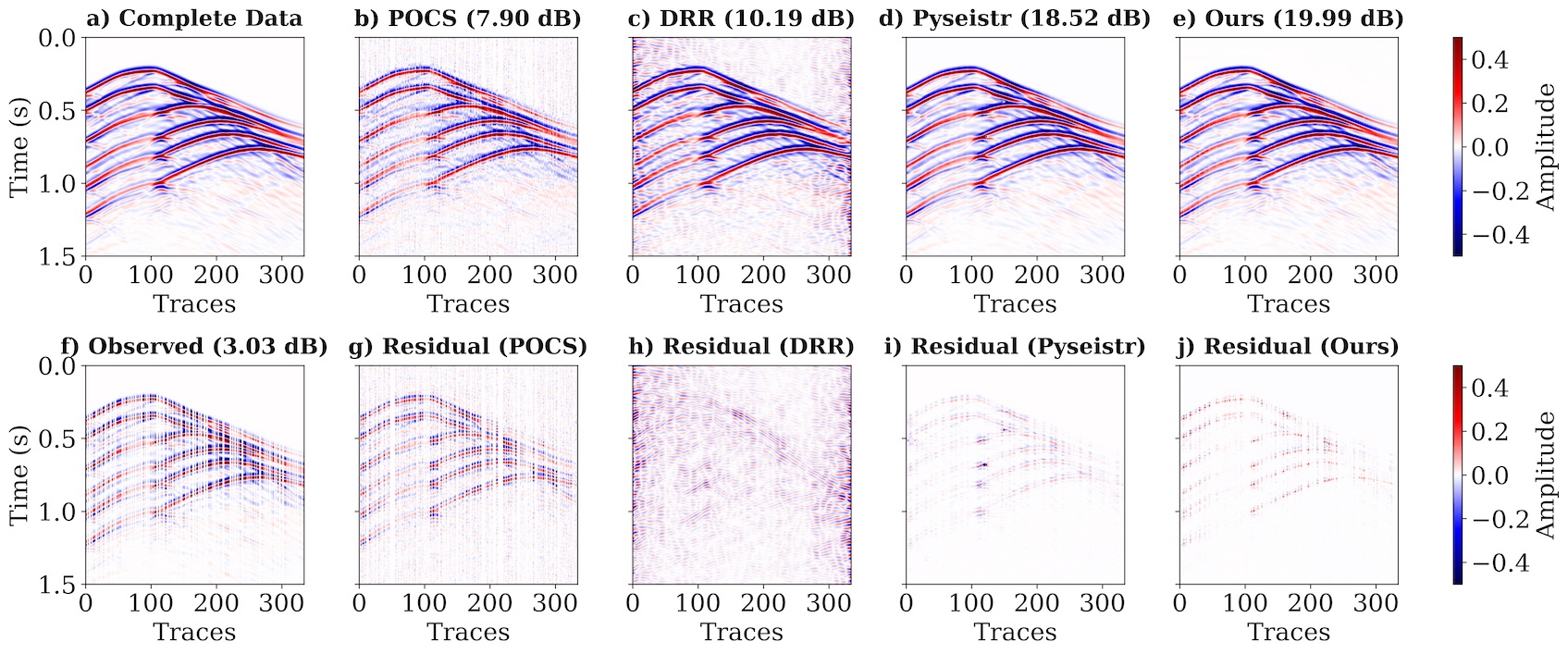}
    \caption{Example 4. Reconstruction of a synthetic DAS gather with 50\% irregular sampling. (a) Complete data. (b)--(e) Reconstruction results of POCS, DRR, PySeisTr, and the proposed method. (f) Observed data with 50\% missing traces. (g)--(j) Corresponding residuals.}
    \label{fig:das4}
\end{figure*}

POCS (Fig.~\ref{fig:das4}b; 7.90\,dB) introduces vertical striping and fails to reconstruct detailed events. DRR (Fig.~\ref{fig:das4}c; 10.19\,dB) improves continuity but leaves incoherent energy in regions with rapid dip changes. PySeisTr (Fig.~\ref{fig:das4}d; 18.52\,dB) preserves reflector geometry more effectively but still exhibits leakage.

The proposed method (Fig.~\ref{fig:das4}e; 19.99\,dB) yields the most stable reconstruction with coherent amplitude variations and minimal artifacts. The residuals (Fig.~\ref{fig:das4}j) confirm accurate DAS-event preservation.

\subsection{Example 5: Viking Graben 2D Field Data}

\begin{figure*}[t]
    \centering
    \includegraphics[width=0.7\textwidth]{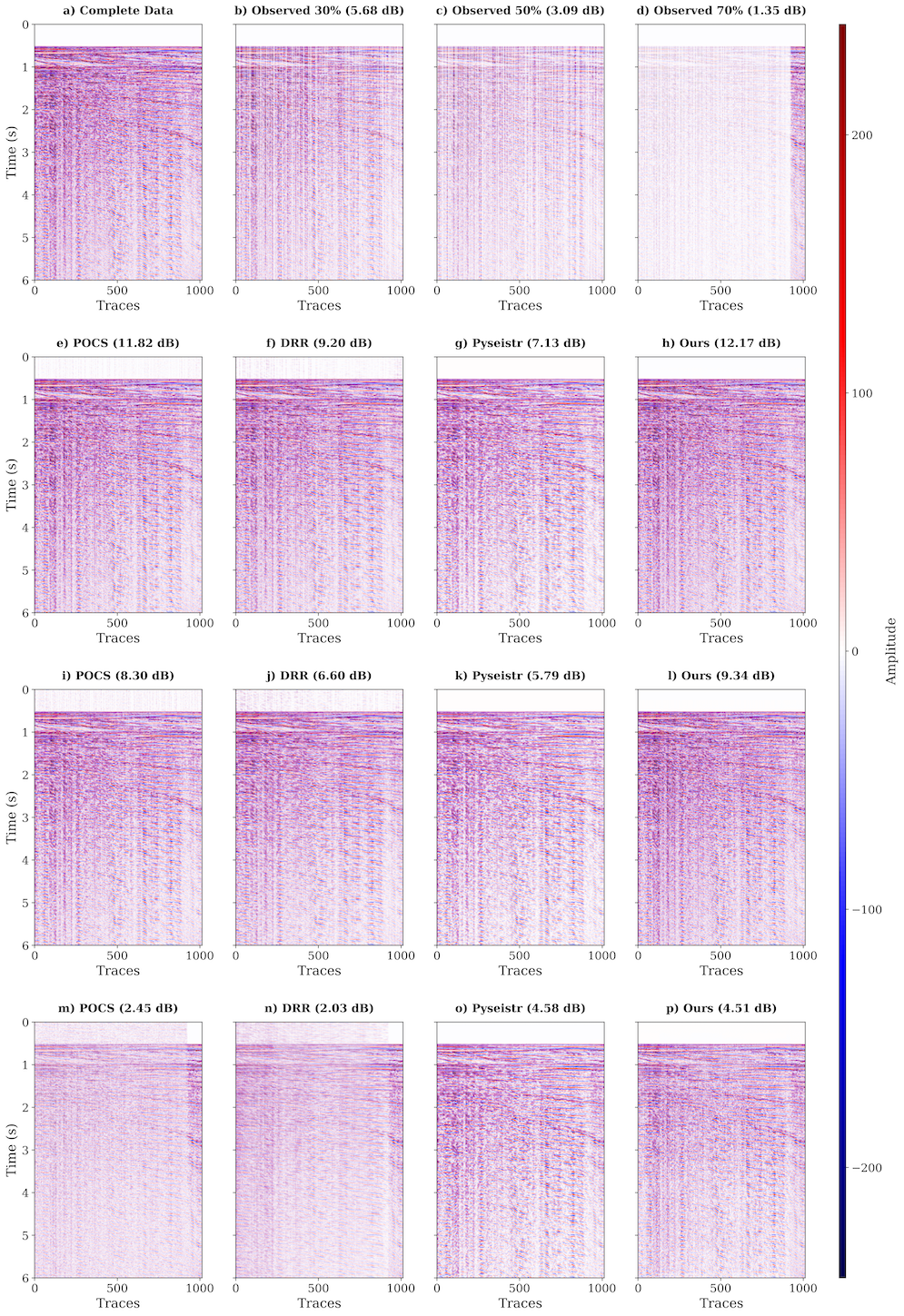}
    \caption{Example 5. Reconstruction of Viking Graben 2D field data with 30\%, 50\%, and 70\% random trace removal. (a) Complete post-stack data. (b)--(d) Observed data. (e)--(p) Reconstructions for all methods at each removal level.}
    \label{fig:viking2d}
\end{figure*}

This experiment uses the well-known Viking Graben 2D poststack section with dimensions $1500 \times 1012$ and a sampling interval of 4\,ms. Randomly removing 30\%, 50\%, and 70\% of the traces provides three reconstruction settings (Figs.~\ref{fig:viking2d}b--d).

For the 30\% removal case, the proposed method achieves the highest accuracy (12.17\,dB), outperforming POCS (11.82\,dB), DRR (9.20\,dB), and PySeisTr (7.13\,dB). Under 50\% removal, the proposed method again leads with 9.34\,dB. The most challenging 70\% case shows severe degradation for all methods, but the proposed model remains competitive at 4.51\,dB with better reflector continuity than the baselines.

\subsection{Example 6: SeanS3 3D Field Data}

\begin{figure*}[t]
    \centering
    \includegraphics[width=0.9\textwidth]{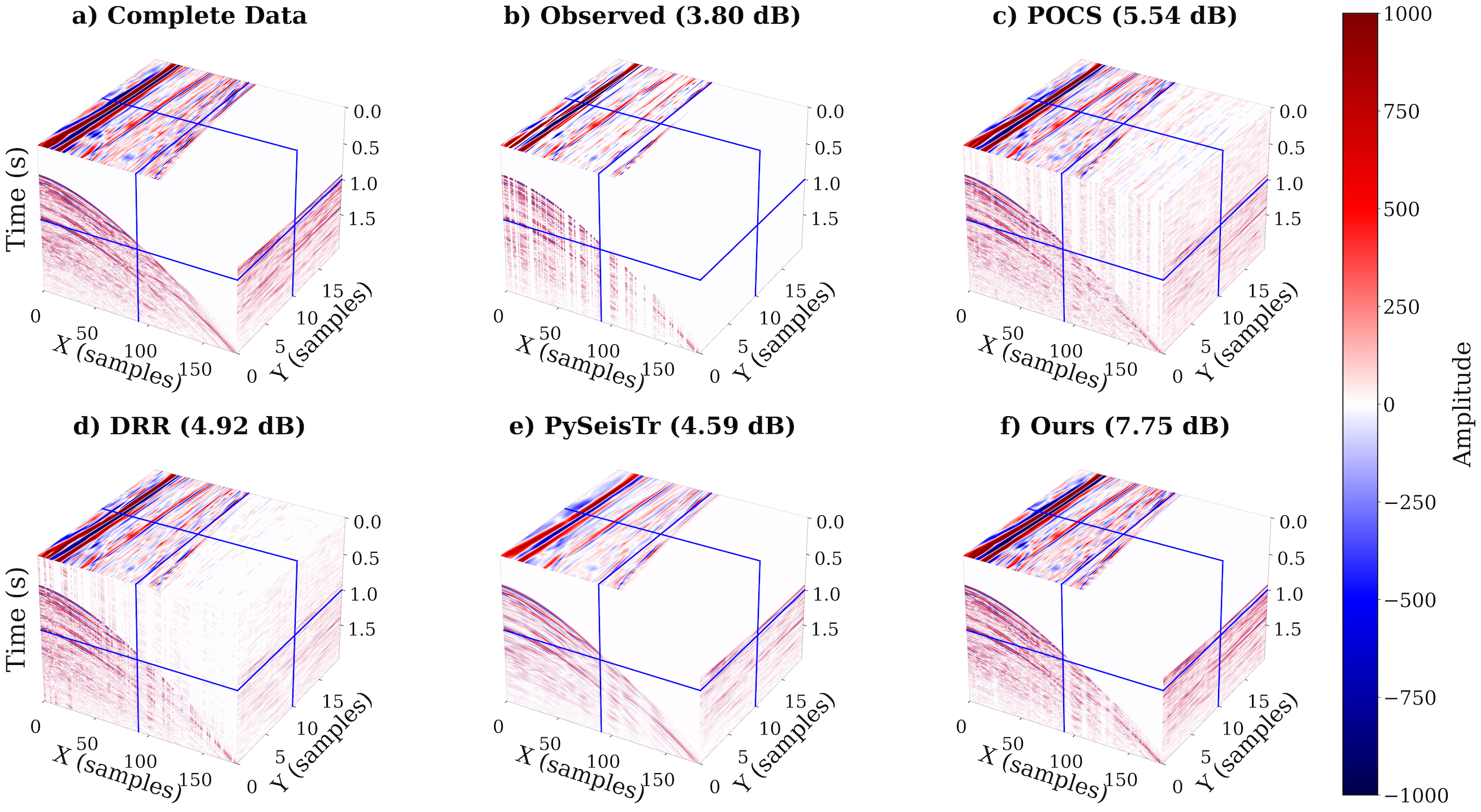}
    \caption{Example 6. Reconstruction of SeanS3 3D field data with 50\% missing traces. (a) Complete volume. (b) Observed data. (c)--(f) Reconstructions from POCS, DRR, PySeisTr, and the proposed model.}
    \label{fig:seans3d}
\end{figure*}

The SeanS3 3D volume contains 500 time samples, 180 receivers, and 20 inlines. With 50\% missing traces (Fig.~\ref{fig:seans3d}b), POCS (5.54\,dB) preserves major events but leaves strong imprinting artifacts. DRR (4.92\,dB) and PySeisTr (4.59\,dB) oversmooth the data. The proposed method achieves 7.75\,dB with clearer event continuity and reduced artifacts.

\subsection{Example 7: Field DAS Section}

\begin{figure*}[t]
    \centering
    \includegraphics[width=0.9\textwidth]{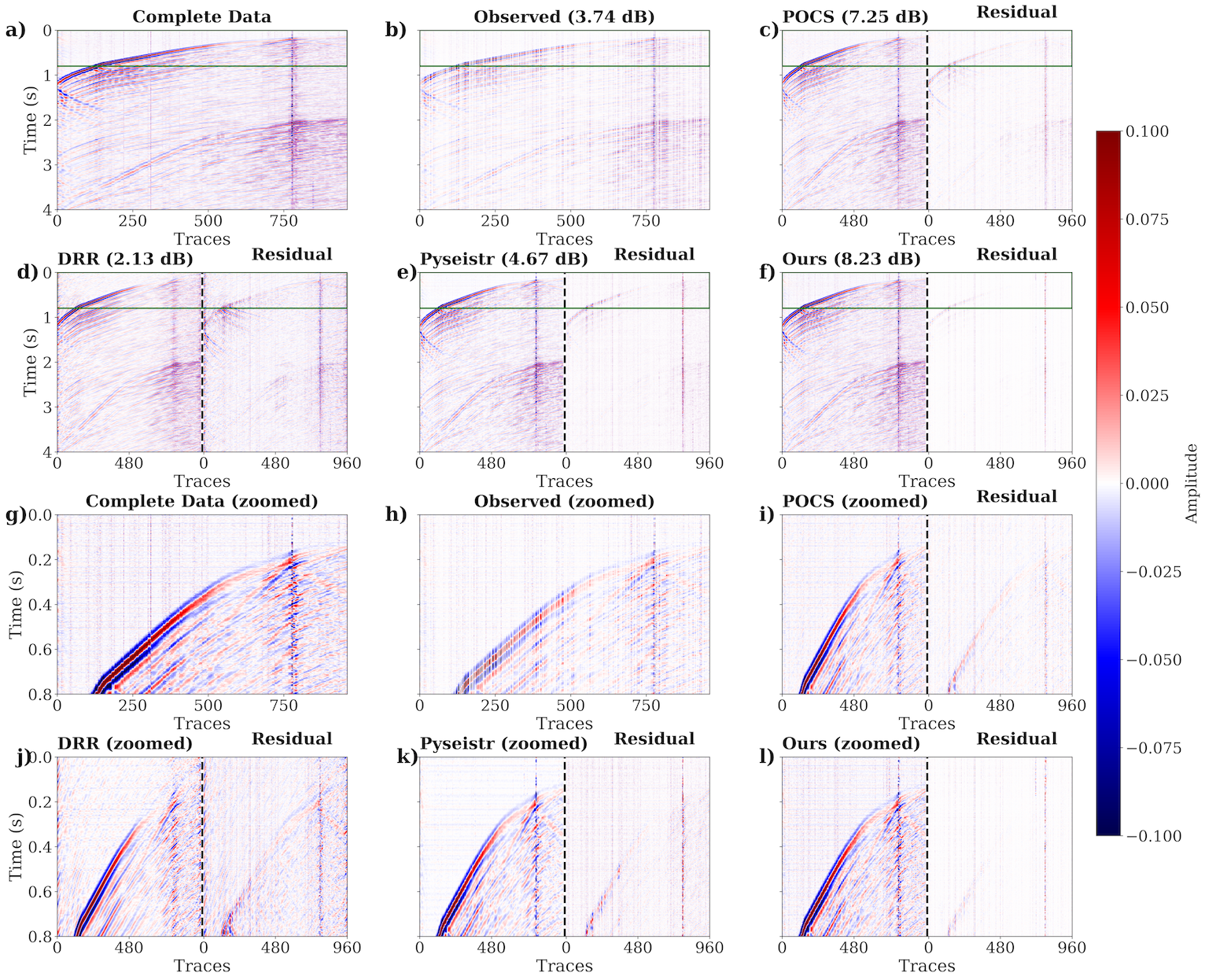}
    \caption{Example 7. Reconstruction of a zoomed field DAS segment with 50\% irregular sampling. (a)--(f) Complete, observed, and reconstructed data for all methods. (g)--(l) Zoomed and residual views.}
    \label{fig:das5}
\end{figure*}

A zoomed region of a field DAS gather from Chen et al.~\citep{Chen2023DAS} is extracted to evaluate reconstruction of fine-scale DAS features. The incomplete input (Fig.~\ref{fig:das5}b; 3.74\,dB) exhibits disrupted curvature and discontinuous events.

POCS (7.25\,dB) restores part of the dominant arrival but introduces striping artifacts. DRR (2.13\,dB) fails to preserve event curvature. PySeisTr (4.67\,dB) produces smoother events but retains leakage.

The proposed method (8.23\,dB) achieves the most coherent event recovery and yields the highest metrics in Table~\ref{tab:das5}, indicating superior preservation of DAS waveform texture.

\begin{table}[t]
\caption{Quantitative comparison for Example~7.}
\centering
\begin{tabular}{lcccc}
\toprule
Method & SNR (dB) & PSNR (dB) & MSE & SSIM \\
\midrule
Observed & 3.74 & 36.65 & $1.1760\times10^{-4}$ & 0.9569 \\
POCS & 7.25 & 40.16 & $5.2355\times10^{-5}$ & 0.9858 \\
DRR & 2.13 & 35.04 & $1.7036\times10^{-4}$ & 0.9403 \\
PySeistr & 4.67 & 37.58 & $9.4969\times10^{-5}$ & 0.9844 \\
Ours & \textbf{8.23} & \textbf{41.14} & \textbf{$4.1839\times10^{-5}$} & \textbf{0.9943} \\
\bottomrule
\end{tabular}
\label{tab:das5}
\end{table}

\subsection{Example 8: Reconstruction and Migration Imaging on the 1997 BP Dataset}

The final experiment assesses how interpolation quality affects migration imaging. We use 385 shot gathers from the 1997 BP model, each containing 256 traces and 384 samples with a sampling interval of 10\,ms. Because some 1997 BP slices are included in SWAN, this is not a blind test; instead it evaluates downstream imaging.

\begin{figure*}[t]
    \centering
    \includegraphics[width=\textwidth]{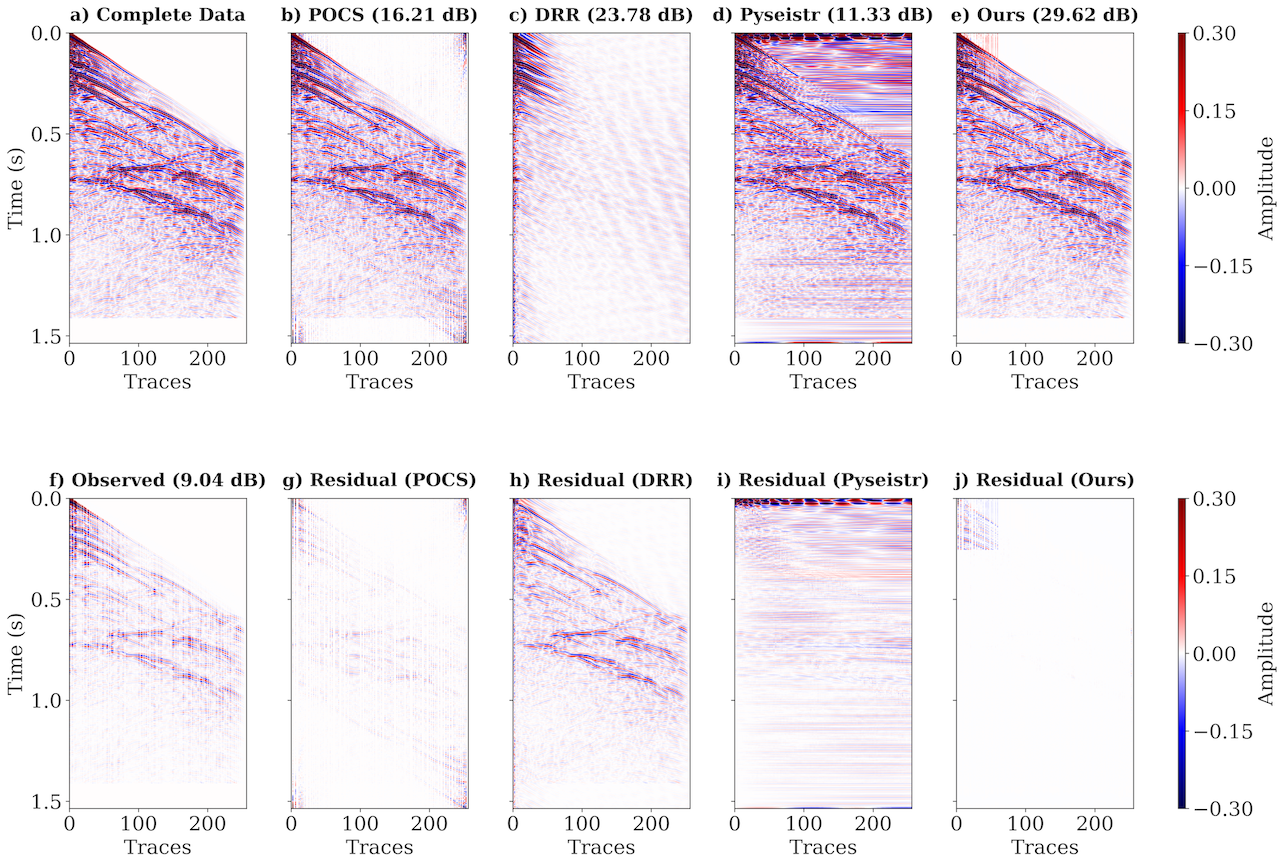}
    \caption{Example 8. Reconstruction of a 1997 BP shot gather with 50\% missing traces.}
    \label{fig:bp_gathers}
\end{figure*}

Reconstruction under 50\% removal is shown in Fig.~\ref{fig:bp_gathers}. POCS (16.21\,dB) partially restores events but introduces vertical streaking. DRR (23.78\,dB) oversmooths dipping energy. PySeisTr (11.33\,dB) reduces noise but smears reflectors. The proposed method achieves 29.62\,dB with coherent dipping events and minimal artifacts.

\begin{figure*}[t]
    \centering
    \includegraphics[width=0.6\textwidth]{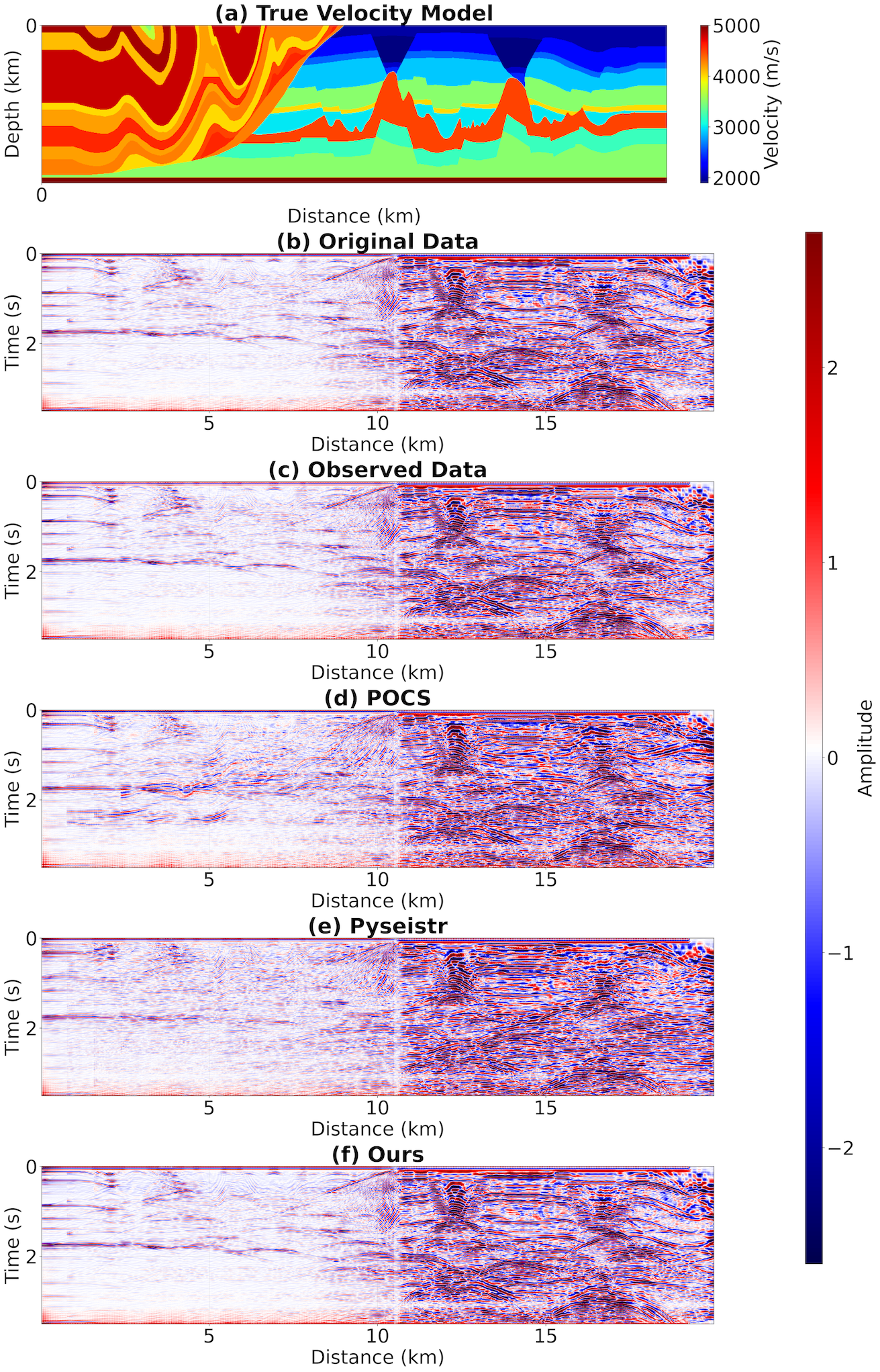}
    \caption{PSTM imaging results for the 1997 BP model.}
    \label{fig:bp_migration}
\end{figure*}

To evaluate imaging quality, we apply CMP sorting, NMO correction, stacking, and Kirchhoff PSTM. The velocity model is converted from SEG-Y to RSF and resampled. Migration of incomplete data results in severe reflector breakage (Fig.~\ref{fig:bp_migration}c). POCS and PySeisTr improve reflector visibility but leave noise and smearing. DRR oversmooths the gathers.

Migration using the proposed reconstruction (Fig.~\ref{fig:bp_migration}g) produces the cleanest section with sharp dips, reduced migration noise, and well-focused reflectors. This experiment highlights how reconstruction quality directly influences seismic imaging.

\section{Discussion}

\subsection{Generalization Enabled by SWAN}

The SWAN dataset is designed to reflect the physical consistency of seismic waveforms across a broad range of geological and acquisition scenarios. It includes hyperbolic reflections from horizontally layered media, dipping and curved events associated with structural deformation, laterally truncated terminations that resemble fault or stratigraphic edges, pseudo three-dimensional reflector continuity across inline and crossline directions, and a wide range of noise characteristics observed in both marine and land surveys. This diversity allows the learning model to acquire a statistically stable prior that captures essential kinematic and dynamic properties of seismic reflections.

The numerical experiments demonstrate that the model consistently reconstructs key seismic attributes in a wide spectrum of settings. The reconstructed gathers preserve expected event geometry, amplitude behavior, and continuity, regardless of whether the input is a simple synthetic hyperbolic pattern, an edge structure with abrupt terminations, a volumetric three-dimensional reflector, or a field dataset such as the Viking Graben section, the SeanS3 volume, or the DAS acquisitions. This stability across datasets indicates that SWAN provides a unified event-level representation that cannot be learned from a single survey and is essential for cross-survey generalization.

A major factor behind this generalization capability lies in the standardized design of the dataset. All wavefields are converted into non-overlapping patches of identical size, normalized consistently, and filtered using a unified quality control rule. Metadata describing sampling intervals, normalization factors, and acquisition characteristics are recorded for each patch. These procedures eliminate survey-specific preprocessing variations that often hinder cross-survey learning in seismic applications. As a result, SWAN offers a reproducible and coherent representation of seismic waveforms that supports large-scale training and robust transfer to new datasets. This design provides a solid foundation for future community benchmarks, where reproducibility and consistent preprocessing are necessary for fair comparison among different reconstruction methods.

\subsection{Deterministic Residual Pathways in RGDM}

The advantages of the proposed RGDM become apparent when considering its residual-guided diffusion mechanism and the deterministic behavior of its latent space evolution. Classical diffusion models operate by gradually corrupting data toward Gaussian noise and learning to reverse this corruption through stochastic sampling steps. Such a formulation is not aligned with seismic degradation patterns, which arise from spatially coherent missing traces rather than random noise. In contrast, RGDM evolves through residual increments that describe the mismatch between the observed data and the underlying clean waveform. This leads to a correction pathway that preserves physically plausible event geometry throughout the diffusion process.

The deterministic nature of the reverse trajectory further reduces sampling variance and avoids the generation of spurious reflections. This property is crucial for seismic applications, where small amplitude variations, subtle edge terminations, and reflector continuity can strongly influence interpretation and downstream imaging. Because SWAN exposes the model to a broad variety of reflection patterns and noise regimes, the latent space learned by RGDM contains a richer representation of seismic waveform characteristics than would be possible with a narrow or single survey training set. Consequently, RGDM follows stable and physically meaningful diffusion paths even when applied to data from geological environments or acquisition geometries not present in the training examples.

\subsection{Limitations and Future Directions}

Although RGDM performs well on the three-dimensional volume considered in this study, the current framework remains limited by its two-dimensional formulation and network. When applied to volumetric data, the model processes each slice independently. True three-dimensional reflector continuity is therefore not explicitly represented, and crossline relationships must be inferred indirectly. As a result, strongly curved horizons, fault-bounded geometries, and other volumetric features may be imperfectly reconstructed when the degree of out-of-plane variation is high. These limitations highlight that both the dataset and the model architecture need to be extended toward explicit three-dimensional representations in order to fully address the complexity of modern seismic acquisitions.

Beyond the reconstruction task, the combination of large-scale datasets and generative models offers promising directions for future research. Diffusion models can be used not only as reconstruction engines, but also as data synthesizers and augmenters that support pretraining, domain adaptation, and cross-survey generalization. With appropriate extensions, SWAN may evolve into a foundation-level training corpus for seismic processing tasks, enabling models that learn transferable representations of waveform physics.

Scalability and reproducibility are central considerations for future development. Expanding SWAN to include additional geological provinces such as deepwater basins, arid land environments, and borehole-oriented surveys would increase the diversity of acquisition conditions. Incorporating wide-azimuth, ocean-bottom-node, and irregular dense sampling would further broaden the range of structural and kinematic patterns available for training. Standardizing metadata and survey descriptors across all components will be essential for establishing a long-term community benchmark that supports reproducible experimentation and fair comparison among different waveform processing methods. Ultimately, these efforts will pave the way toward next-generation foundation models for seismic data processing and imaging.

\section{Conclusion}

We introduced SWAN, a large-scale open-source seismic waveforms dataset designed to support the development of generalizable deep-learning models for seismic processing. Building upon this dataset, we proposed the Residual-Guided Diffusion Model (RGDM), which reformulates diffusion as a deterministic residual-correction process anchored to the observed waveform. RGDM leverages the structured nature of seismic residuals and the rich event-level priors learned from SWAN to achieve accurate and physically consistent reconstruction with only a few diffusion steps. Extensive experiments on synthetic, pseudo-3D, and field data (the Viking Graben 2D section and SeanS3 3D volume) demonstrate that RGDM outperforms established baselines, including POCS, DRR, and PySeistr, particularly in scenarios with severe missing-trace or complex reflector features. These results highlight the importance of combining diverse training data with diffusion models tailored to seismic physics. The SWAN dataset and RGDM framework together provide a robust foundation for future research in machine-learning-based seismic processing, and we anticipate that they will facilitate further advances in interpolation, denoising, imaging, and multi-dimensional seismic reconstruction.

\section{Data Availability Statement}

The datasets will be available in the public domain.

\bibliographystyle{plainnat}
\bibliography{swref}

\end{document}